\documentclass[12pt,preprint]{aastex}

\begin{document}


\title{Detection of a $63\arcdeg$ Cold Stellar Stream in the Sloan Digital Sky Survey}

\author{C. J. Grillmair}
\affil{Spitzer Science Center, 1200 E. California Blvd., Pasadena,  CA 91125}
\email{carl@ipac.caltech.edu}
\and
\author{O. Dionatos\altaffilmark{1}}
\affil{INAF - Osservatorio Astronomico di Roma, Via di Frascati 33,
00040, Monteporzio Catone, Italy}

\email{dionatos@mporzio.astro.it}

\altaffiltext{1}{Department of Physics, University of Athens,
Panepistimiopolis, 15771 Athens, Greece}

\begin{abstract}

We report on the detection in Sloan Digital Sky Survey data of a
$63\arcdeg$-long tidal stream of stars, extending from Ursa Major to
Cancer. The stream has no obvious association with the orbit of any
known cluster or galaxy. The contrast of the detected stream is
greatest when using a star count filter that is matched to the
color-magnitude distribution of stars in M 13, which suggests that the
stars making up the stream are old and metal poor. The visible portion
of the stream is very narrow and about 8.5 kpc above the Galactic
disk, suggesting that the progenitor is or was a globular
cluster. While the surface density of the stream varies considerably
along its length, its path on the sky is very smooth and uniform,
showing no evidence of perturbations by large mass concentrations in
the nearby halo. While definitive constraints cannot be established
without radial velocity information, the stream's projected path and
estimates of its distance suggest that we are observing the stream
near the perigalacticon of its orbit.

\end{abstract}



\section{Introduction}

Despite its still limited extent, the Sloan Digital Sky Survey (SDSS)
continues to be a remarkable resource for studies of Galactic
structure. In addition to the large scale features attributed to past
galaxy accretion events \citep{yann03,maje2003,roch04}, SDSS data were
used to detect the remarkably strong tidal tails of Palomar 5
\citep{oden2001,rock2002, oden2003}. Tidal tails of globular clusters
are particularly interesting from a Galactic structure standpoint 
as they are expected to be very numerous and to sample the Galactic
potential much more uniformly than satellite galaxies. Moreover, such
tidal tails are dynamically very cold \citep{comb99},
making them useful for constraining not only the global Galactic
potential, but also its lumpiness \citep{mura99}.

Substantial tidal streams have now been found associated with at least
two of the eight globular clusters in the SDSS area; Pal 5
\citep{oden2003, grill2006b} and NGC 5466 \citep{belo05,
grill2006a}. In this Letter we examine a much larger region of the SDSS
to search for more extended structures. We briefly describe our
analysis in Section \ref{analysis}. We discuss the detection of a new
stellar stream in Section \ref{discussion}, make some initial distance
estimates in Section \ref{distance}, attempt to identify a progenitor
in Section \ref{progenitor}, and put initial constraints on the orbit
in Section \ref{orbit}. We make concluding remarks Section
\ref{conclusions}.

\section{Data Analysis \label{analysis}}

Data comprising $u',g',r',i'$, and $z'$ photometry for $5.3 \times
10^7$ stars in the region $124\arcdeg < \alpha < 251\arcdeg$ and
$-1\arcdeg < \delta < 65\arcdeg$ were extracted from the SDSS database
using the SDSS CasJobs query system. The data were analyzed using the
matched filter technique employed by \citet{grill2006a} and
\citet{grill2006b}, and described in detail by \citet{rock2002}. This
technique is made necessary by the fact that, over the magnitude range
and over the region of sky we are considering, the foreground disk
stars outnumber the more distant stars in the Galactic halo by three
orders of magnitude. Applied to the color-magnitude domain, the
matched filter is a means by which we can optimally differentiate
between two populations, provided we know the color-magnitude
distribution of each.

In practise, we use the SDSS photometry to create a color-magnitude
density or Hess diagram for both the stars of interest (e.g. those in
a globular cluster) and the foreground population. Dividing the
former by the latter, we generate an array of relative weights that 
constitute an optimal color-magnitude filter.  Using this filter,
every star in the survey can be assigned a weight or probability of
being a member of our chosen globular cluster based on its measured
magnitude and color. Having used observed data to generate it, the
filter implicitly includes the effects of photometric uncertainties,
and even though a particular star may lie 3$\sigma$ away from the main
sequence of a globular cluster of interest, its relative weight
(and thus probability of being a cluster star) may still be high if
the number of foreground stars in this region of the Hess diagram is
relatively low (e.g. near the horizontal branch, or blueward of the
main sequence turnoff).

As we were initially interested in searching for tidal streams
associated with globular clusters, we constructed Hess diagrams for
each of the eight globular clusters in the SDSS DR4 area (NGC 2419,
Pal 3, NGC 5272, NGC 5466, Pal 5, NGC 6205, NGC 7078, and NGC 7089).
A single Hess diagram for field stars was generated using $1.2 \times
10^7$ stars spread over $\sim 2200$ deg$^2$ of DR4.  We then
applied each of the eight resulting optimal filters in turn to the entire
survey area. The resulting weighted star counts were summed by
location on the sky to produce eight different, two dimensional weight
images or probability maps. 

We used all stars with $15 < g' < 22.5$. We dereddended the SDSS
photometry as a function of position on the sky using the DIRBE/IRAS
dust maps of \citet{schleg98}. We optimally filtered the $g' - u'$,
$g' - r'$, $g' - i'$, and $g' - z'$ star counts independently and then
co-added the resulting weight images.  In Figure 1 we show the final,
combined, filtered star count distribution, using a filter matched to
the color magnitude distribution of stars in NGC 6205 (M 13). The
image has been smoothed with a Gaussian kernel with $\sigma =
0.2\arcdeg$. A low-order, polynomial surface has been subtracted from
the image to approximately remove large scale gradients due to the
Galactic disk and bulge.

\section{Discussion \label{discussion}}

Quite obvious in Figure 1 is a long, remarkably smooth, curving stream
of stars, extending over $140\arcdeg < \alpha < 220\arcdeg$. On
the sky the stream runs in an almost straight line through the whole
of Ursa Major and Leo Minor, ending in Cancer and spanning a total of
$63 \arcdeg$. The stream is most evident when we use a filter that is
matched to the color-magnitude distribution and luminosity function of
stars in M 13, although shifted faintwards by 0.2 mag.
Optimal filters based on the other seven globular clusters in DR4 did
not yield the level of contrast that we see in Figure 1. The stream is
easily visible in each individual color pair, including $g' -
u'$. There may be a second, more diffuse feature with $174 < \alpha < 200$ 
about $3\arcdeg$ to the north of the stream, but we defer analysis 
of this feature to  a future paper.

The stream is not a product of our dereddening procedure; careful
examination of the reddening map of \citet{schleg98} shows no
correlation between this feature and the applied reddening
corrections. The maximum values of $E(B-V)$ are $\approx 0.03$, with
typical values in the range 0.01 - 0.02 over the length of the
stream.  Rerunning the matched filter analysis without reddening
corrections yielded little more than a slight reduction in the
apparent strength of the stream.

We also ran our optimum filter against the SDSS DR4 galaxy
catalog to investigate whether the stream could be due to confusion
with faint galaxies. (Such a structure in the distribution of galaxies
would be no less interesting than a stellar stream of these
dimensions!). However, we found no feature in the filtered galaxy counts
that could mimic the stream apparent in Figure 1.

At its southwestern end the stream is truncated by the limits of the
available data. We attempted to trace the stream in the portion of DR4
with $0\arcdeg < \delta < 10\arcdeg$ but could find no
convincing continuation. Plausible orbits for the stream (see below)
predict a fairly narrow range of possible paths across this region,
and generally a rather sharp increase in Sun-stream distance.  We
attempted to recover the stream by shifting our filter from -1.0 to
+3.0 mags, but to no avail. A continuation of the stream may well be
there, but the power of the matched filter is significantly reduced as
the bulk of the main sequence drops below the survey data's 50\%
completeness threshold. Combined with the rapid rise in the number of
contaminating Galactic disk stars in this region, there appears to be
little chance of recovering the stream until much deeper surveys
become available.

On the northeastern end, the stream becomes indiscernible beyond RA =
$220\arcdeg$.  We attempted to enhance the northeastern end of the
stream by shifting the filter to both brighter and fainter magnitudes,
but again to no avail. Experiments in which we inserted artificial
stellar streams with surface densities similar to those observed in
Figure 1 revealed that they too largely vanished beyond R.A. =
$220\arcdeg$. Hence, while it is conceivable that we are seeing the
physical end of the stream, it is equally possible that our failure to
trace the stream any further reflects once again the rapidly
increasing contamination by field stars at lower Galactic latitudes.

Sampling at several representative points, the stream appears to be
$30 \arcmin$ wide (FWHM) on average. This width is similar to those
observed in the tidal tails of the globular clusters Pal 5 and NGC
5466 \citep{grill2006b,grill2006a}. On the other hand, the width is
much narrower than the tidal arms of the Sagittarius dwarf
\citep{maje2003,mart2004} (one of which runs along the southern edge
of the field shown in Figure 1). This suggests that the stars making
up the stream have a low random velocities, and that they were
probably weakly stripped from a relatively small potential. Combining
this with a location high above the Galactic plane (see below)
suggests that the parent body is or was a globular cluster.

Integrating the background subtracted, weighted star counts over a
width of $\approx 0.8\arcdeg$ we find the total number of stars in the
discernible stream to be $1800 \pm 200$. As is evident in Figure 1,
the surface density of stars fluctuates considerably along the
stream. For stars with $g' < 22.5$ the average surface density is $25
\pm 5$ stars deg$^{-2}$, with occasional peaks of over 70 stars
deg$^{-2}$.

\subsection{Distance to the Stream \label{distance}}

The power of the matched filter resides primarily at the main sequence
turnoff and below, where the luminosity function increases rapidly and
the stars lie blueward of the bulk of foreground population.  The blue
horizontal branch can generate much higher weights per star, but the
typical numbers of horizontal branch stars in any likely progenitor
are too low to account for such a continuous and well-populated stream
(e.g. \citet{grill2006a}).  Assuming that our filter is indeed beating
against the main sequence of the stream population, we can use the
filter response to estimate distances. We have attempted to extract
the color magnitude distribution for the stream stars directly, but
contamination by foreground stars is so high as to make
differentiation between stream and field distributions highly
uncertain.

Varying the shift applied to the M 13 matched filter from -0.3 to +0.7
mags, we measured the background-subtracted, mean weighted star counts
along the stream in the regions $140\arcdeg < \alpha < 154\arcdeg$,
$154\arcdeg < \alpha < 180\arcdeg$, and $180\arcdeg < \alpha <
220\arcdeg$.  We also measured a $1\arcdeg$ segment centered on the
strongest concentration of stream stars at (R.A, decl.) =
(144.1\arcdeg, 30.3\arcdeg). To avoid potential problems related to a
difference in age between M 13 and the stream stars, we used only the
portion of the filter with $19.5 < g' < 22.5$, where the bright cutoff
is 0.8 mags below M 13's main sequence turn off. This reduces the
contrast between the stream and the background by about 40\%, but
still provides sufficient signal strength to enable a reasonably
precise measurement of peak response. If our assumptions above are valid,
we are effectively measuring relative distances by main sequence
fitting.

The mean stream surface densities as a function of filter magnitude
shift are shown in Figure 2. Using a distance to M 13 of 7.7 kpc and
fitting Gaussians to the individual profiles, we find Sun-stream
distances of 7.3, 8.5, and 9.1 kpc, respectively, for the three stream
segments identified above. The high density clump at R.A. $=
144\arcdeg$ yields a distance of 7.7 kpc. This puts the mean distance
of the stream high above the Galactic disk at $\sim 8.5$ kpc, with the
stream oriented almost perpendicularly to our line of sight.  The
corresponding Galactocentric distances range from 13.5 to 15 kpc.  The
measured distances rise more or less monotonically from southwest to
northeast, as might be expected for a feature that traces a small part
of an extended Galactic orbit. Based on the widths of the peaks in
Figure 2 we estimate our random measurement uncertainties to be
$\approx 500$ pc.

While the match between the color-magnitude distributions of stars in
M 13 and those in the stream is uncertain, the relative line-of-sight
distances along the stream should be fairly robust. The measurement of
relative distances rests only on the assumption that the
color-magnitude distribution of stream stars is uniform, and that
different parts of the stream will respond to filter shifts in the
same way. Of course, this may not be valid if the luminosity function
of stars escaping from the parent cluster changed as a function of
time due to the dynamical evolution of the cluster, and a forthcoming
paper will deal with possible observational support for this. Our
relative distance estimates may also be subject to variations in SDSS
sensitivity and completeness at faint magnitudes, although it seems
reasonable to suppose that such variations will have largely averaged
out on the scales with which we are dealing.

\subsection{The Stream Progenitor \label{progenitor}}

The location and the narrowness of the stream lead us to believe that
the progenitor likely is or was a globular cluster. On the other hand,
the stream does not pass near any of the eight globular clusters in
DR4. In particular, despite the apparent similarity in color magnitude
distribution, there appears to be no way to dynamically associate the
stream with M 13. The distance, radial velocity, and proper motion
measurements for M 13 are among the best for any globular cluster
\citep{oden97,dine99}. Using these measurements, the projected orbital
path of M 13 is neither near nor aligned with the stream for either
the last two or the next two orbits of the cluster. We conclude that
there is almost no possibility that M 13 could be the progenitor of
the current stream.

Could it be that the stream might be an apogalactic concentration of
tidally stripped stars from a globular cluster that is currently
elsewhere in its orbit? We have integrated orbits for 39 globular
clusters with measured proper motions
\citep{oden97,dine99,chen2000,dine2000,dine2001,sieg2001,wang2000} to
see whether their orbits are aligned with the stream at its current
location. This is fraught with considerable uncertainty as even small
errors in the proper motions measurements can lead to rather
large departures between the predicted and true orbits of clusters.
Nonetheless, as a first attempt to associate this stream with a
progenitor, we integrated orbits for the 39 clusters, projected them
onto the sky, and compared them with the location and orientation of
the stream. We used the Galactic model of \citet{allen91}, which
includes a disk and bulge and assumes a spherical halo.

Only two of the 39 clusters, NGC 1904 and NGC 4590, have predicted
orbits whose projections lie near the stream on the sky and are
roughly aligned with it. Both of these clusters show evidence of tidal
tails \citep{grill95}. However, for NGC 1904 the orbital path in the
region of the stream passes within 5 kpc of the Sun.  This would
require a brightward shift of the M 13 filter of over a magnitude, a
shift that we have tested and for which the stream all but
disappears. For NGC 4590, the apparent alignment with the stream is
much closer, but the predicted distances at the eastern and
western ends of the stream are 8.2 and 19 kpc, respectively. Thus, not
only does the mean distance disagree with our estimate above, but the
spatial orientation of the stream is at odds with the predicted orbit
by more than $50\arcdeg$.

Where then is the progenitor of the stream?  It is certainly possible
that the stream represents the leavings of one of the $\sim 108$
clusters for which we cannot yet estimate orbits.  Alternatively, the
source of the stream may be embedded in the stream itself. The densest
portions of the stream occur at (R.A., decl.) = ($144.13\arcdeg,
30.3\arcdeg$) and ($157.5\arcdeg, 43.7\arcdeg$), each with surface
densities of over 70 stars deg$^{-2}$.  However, 70 stars deg$^{-2}$
to almost 4 mags below the turn off would barely qualify as an open
cluster. Examining the distribution of SDSS catalog stars directly,
there appears to be no tendency for the stars to cluster in these
regions. The highest density peaks in Figure 1 are probably not bound
clusters, and may rather be entirely analogous to similar peaks seen
in the tidal stream of Pal 5, which are interpreted as a natural
consequence of the episodic nature of tidal stripping.  Alternatively,
one of the peaks in Figure 1 could be the {\it remnant} of the parent
cluster. The existence of such disrupted clusters would not be
unexpected \citep{gned97}, and Pal 5 itself is believed to have lost
at least 90\% of its mass and to be on its last orbit around the
Galaxy \citep{grill2001, oden2003}.

\subsection{Constraints on the Orbit of Stream} \label{orbit}

Lacking velocity measurements, we cannot ``solve'' for the
orbit of the stream. However, the apparent orientation of the stream,
along with our estimates of its distance and curvature, can yield some
constraints. Again using the Galactic model of \citet{allen91} (which
\citet{grill2006a} and \citet{grill2006b} found to work reasonably
well for NGC 5466 and Pal 5), we use a least squares method to fit
both the orientation on the sky and the distance measurements in
Section \ref{distance}.  The tangential velocity at each point is
primarily constrained by the projected path of the stream on the sky,
while our distance estimates help to limit the range of possible
radial velocities.  We fit to a number of normal points lying along
the centerline of the stream and adopt a radial velocity fiducial
point at the midpoint of the northeastern segment above, with R.A.,
dec = (202.0\arcdeg, 58.4\arcdeg).

If we give no weight to our relative distance estimates but use only
the $9.1 \pm 0.5$ kpc estimate for the fiducial point, then for a
reasonable range of radial velocities ($-320$ km s$^{-1} < v_{LSR} < 320$ km
s$^{-1}$), the perigalacticon of the stream's orbit must lie in the
range 6.5 kpc $< R < 13.5$ kpc.  For all LSR radial velocities in the
range -270 km s$^{-1} < v_{LSR} < 0$ km s$^{-1}$, the apogalactic radius is
constrained to fall within 17 kpc $< R < 40$ kpc. On the other hand,
positive radial velocities lead to orbits with apogalactica rising
from 100 to 500 kpc. All of these orbits give excellent fits to the
observed path of the stream on the sky.

If we constrain the model using our relative distance estimates
(allowing the proper motions to become free ranging and uninteresting
parameters) we find a best fit value for the radial velocity at the
fiducial point of $-208 \pm 30$ km s$^{-1}$, where the uncertainty
corresponds to the 95\% confidence interval.  The orbit corresponding
to this radial velocity has $R_p = 13.2 \pm 0.7$ kpc and $R_a = 18 \pm
2$ kpc, where the uncertainties reflect only our estimated uncertainty
in the radial velocity. For this orbit, the physical length of the
visible stream would be $\simeq 9.6$ kpc.

\section{Conclusions \label{conclusions}}

Applying optimal contrast filtering techniques to SDSS data, we have
detected a stream of stars some $63\arcdeg$ long on the sky. We are
unable to identify a progenitor for this stream, although from its
appearance and location on the sky, we believe it to be a either an
extant or disrupted globular cluster. Based on a good match to the
color-magnitude distribution of stars in M 13, we conclude that the
stars making up the stream are primarily old and metal poor, and that
the stream as a whole is about 8.5 kpc distant and roughly
perpendicular to our line of sight.

Refinement of the stream's orbit will require radial velocity
measurements of individual stars along its length.  Ultimately, the
vetted stream stars will become prime targets for the {\it Space
Interferometry Mission}, whose proper motion measurements will enable
very much stronger constraints to be placed on both the orbit of the
progenitor and on the potential field of the Galaxy.

\acknowledgments

We are grateful to Helio Rocha-Pinto for comments which significantly
improved the manuscript. Funding for the creation and distribution of
the SDSS Archive has been provided by the Alfred P. Sloan Foundation,
the Participating Institutions, the National Aeronautics and Space
Administration, the National Science Foundation, the U.S. Department
of Energy, the Japanese Monbukagakusho, and the Max Planck Society.

{\it Facilities:} \facility{Sloan}.

\clearpage



\begin{figure}
\epsscale{1.0}
\plotone{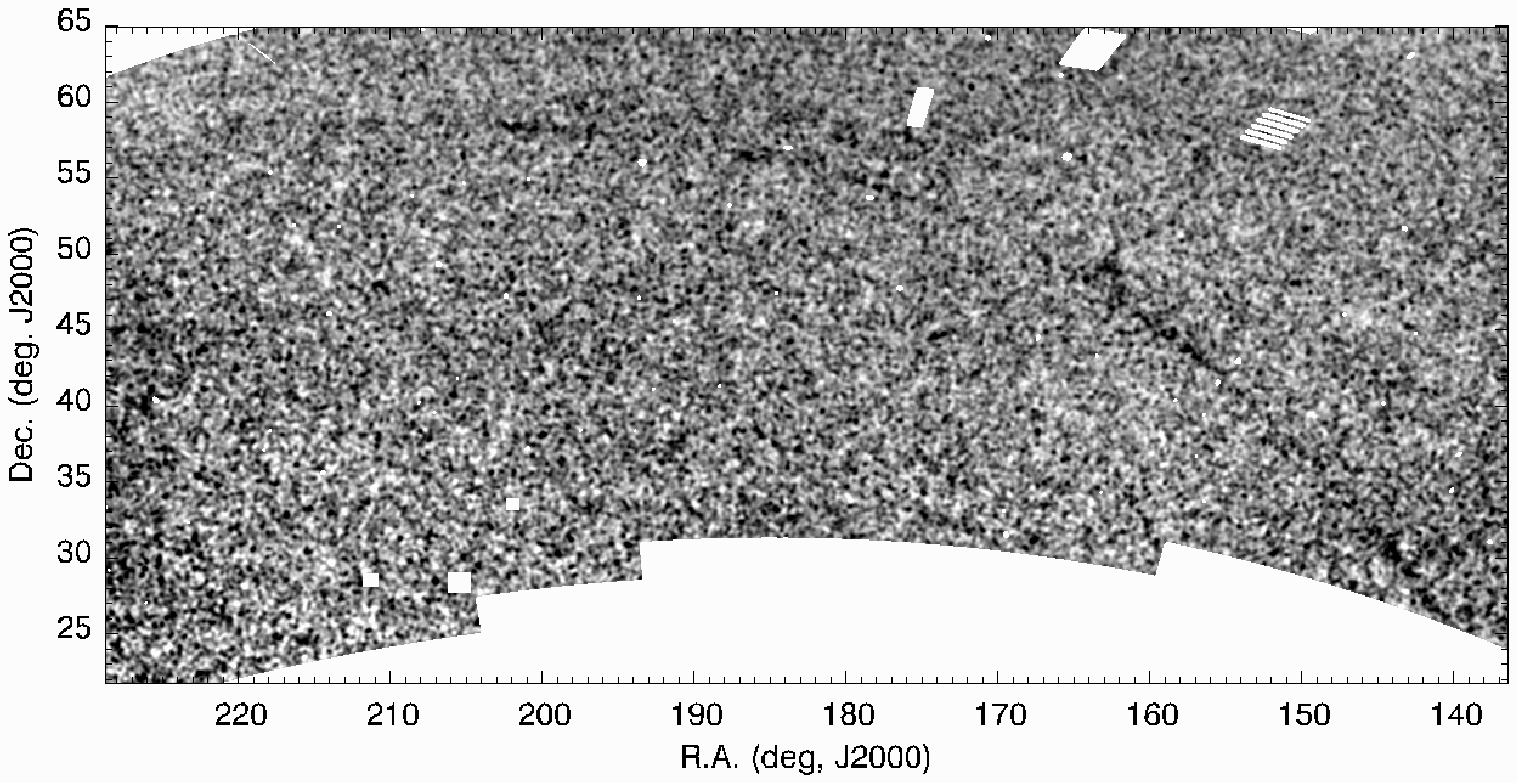}
\caption{Smoothed, summed weight image of the SDSS field after
subtraction of a low-order polynomial surface fit. Darker areas
indicate higher surface densities. The weight image has been smoothed
with a Gaussian kernel with $\sigma = 0.2\arcdeg$. The white areas are
either missing data, or clusters, or bright stars which have been
masked out prior to analysis.\label{fig1}}
\end{figure}

\begin{figure}
\epsscale{0.8}
\plotone{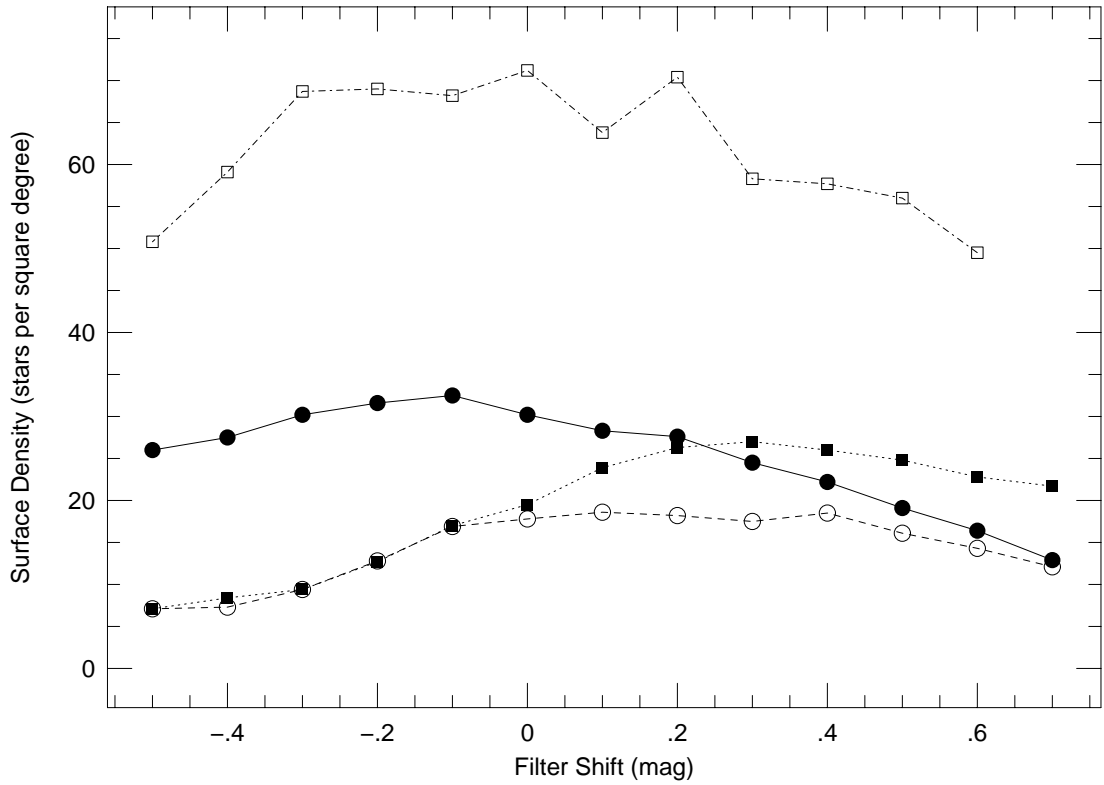}
\caption{The background subtracted, mean surface densities of stream
stars as a function of the magnitude offset applied to the star count
filter. Filled circles correspond to the stream segment spanning
$140\arcdeg < \alpha < 154\arcdeg$, open circles indicate surface
densities for the region $154\arcdeg < \alpha < 180\arcdeg$, and
filled squares show the results for $180\arcdeg < \alpha <
220\arcdeg$. The open squares were determined for a 1 deg$^2$ area
centered on the concentration of stars at R.A., dec = ($144.1\arcdeg,
+30.3\arcdeg$).}
\end{figure}

\end{document}